\documentclass[aps,onecolumn,groupedaddress,showpacs,notitlepage,nofootinbib]{revtex4-1}

\usepackage{graphicx}
\usepackage{dcolumn}
\usepackage{bm}
\usepackage{amsmath}
\usepackage{amsthm}
\usepackage{multirow}
\usepackage{enumerate}

\usepackage{amsfonts}
\usepackage{euscript}
\usepackage{ifthen}
\usepackage{psfrag}
\usepackage{slashed}
\usepackage{hyperref}
\usepackage{gensymb}
\usepackage{color}

\allowdisplaybreaks[4]

\newcommand{\be}{\begin{equation}}
\newcommand{\ee}{\end{equation}}
\newcommand{\bea}{\begin{eqnarray}}
\newcommand{\eea}{\end{eqnarray}}

\begin{document} 
\begin{center}
\hfill MI-TH-1879
\end{center}
\title{21cm Limits on Decaying Dark Matter and Primordial Black Holes}

\author{Steven Clark$^{1}$}
\email{cla07003@physics.tamu.edu}
\author{Bhaskar Dutta$^{1}$}
\email{dutta@physics.tamu.edu}
\author{Yu Gao$^{2}$}
\email{gaoyu@ihep.ac.cn}
\author{Yin-Zhe Ma$^{3,4}$}
\email{ma@ukzn.ac.za}
\author{Louis E. Strigari$^{1}$}
\email{strigari@physics.tamu.edu}

\affiliation{
$^{1}$~Department of Physics and Astronomy, Mitchell Institute for Fundamental Physics and Astronomy, Texas A\&M University, College Station, TX 77843-4242, USA\\
$^{2}$~Key Laboratory of Particle Astrophysics, Institute of High Energy Physics,\\
Chinese Academy of Sciences, Beijing 100049, China\\
$^{3}$School of Chemistry and Physics, University of KwaZulu-Natal,
Westville Campus, Private Bag X54001, Durban, 4000, South Africa\\
$^{4}$ NAOC--UKZN Computational Astrophysics Centre (NUCAC),
University of KwaZulu-Natal, Durban, 4000, South Africa
}

\begin{abstract}
Recently the Experiment to Detect the Global Epoch of Reionization Signature (EDGES) reported the detection of a 21cm absorption signal stronger than astrophysical expectations. In this paper we study the impact of radiation from dark matter (DM) decay and primordial black holes (PBH) on the 21cm radiation temperature in the reionization epoch, and impose a constraint on the decaying dark matter and PBH energy injection in the intergalactic medium, which can heat up neutral hydrogen gas and weaken the 21cm absorption signal. We consider decay channels DM$\rightarrow e^+e^-, \gamma\gamma$, $\mu^+\mu^-$, $b\bar{b}$ and the $10^{15-17}$g mass range for primordial black holes, and require the heating of the neutral hydrogen does not negate the 21cm absorption signal.  For $e^+e^-$, $\gamma\gamma$ final states  and PBH cases we find strong 21cm bounds that can be more stringent than the current extragalactic diffuse photon bounds. For the DM$\rightarrow e^+e^-$ channel, the lifetime bound is $\tau_{\rm DM}> 10^{27}$s for sub-GeV dark matter. The bound is $\tau_{\rm DM}\ge 10^{26}$s for sub-GeV DM$\rightarrow \gamma\gamma$ channel and reaches $10^{27}$s at MeV DM mass. For $b\bar{b}$ and $\mu^+\mu^-$ cases, the 21 cm constraint is better than all the existing constraints for $m_{\rm DM}<20$ GeV where the bound on $\tau_{\rm DM}\ge10^{26}$s. For both DM decay and primordial black hole cases, the 21cm bounds significantly improve over the CMB damping limits from {\it Planck} data.
\end{abstract}

\maketitle



\section{Introduction}
\label{sect:intro}

Recently the Experiment to Detect the Global Epoch of Reionization Signature (EDGES) reported the observation of 21 centimeter absorption lines at highly-redshift $z$=15-20~\cite{Bowman:2018yin}, with a best-fit neutral Hydrogen  spin temperature $T_{\rm S}$ much lower than conventional astrophysical expectations, leading to strong absorption signals~\cite{Bowman:2018yin}.  The abrupt lowering of $T_{\rm S}$ relative to the cosmic microwave background temperature $T_{\rm CMB}$ near $z\sim 20$ has been interpreted~\cite{Bowman:2018yin, Barkana:2018lgd} as due to the re-coupling of $T_{\rm S}$ to the Hydrogen gas temperature $T_{\rm{G}}$ by the Wouthuysen-Field effect~\cite{wouthuysen21cm,field21cm}, where a lower-than-standard Hydrogen gas temperature has been proposed that may arise from cooling effects~\cite{Barkana:2018lgd} via interaction with hypothetical particles. Fulfilling such a role, potential interactions between baryons and cold dark matter (DM) have been studied and constrained~\cite{Fialkov:2018xre, Berlin:2018sjs, Barkana:2018qrx, Fraser:2018acy, Kang:2018qhi,Jin:2016kio}. 

From another perspective, the observation of 21cm signal also places a bound on hypothetical processes that are capable of heating up the intergalactic medium prior to the reionization time, e.g., by the energy injection from the annihilation of dark matter~\cite{DAmico:2018sxd}. In this paper we investigate such a bound on the decay of dark matter and the Hawking radiation~\cite{Hawking:1974rv} from primordial black holes, as both processes emit high-energy electrons and photons throughout the post-recombination history of the Universe.

Beside mapping the Universe's mass distribution at high redshift, the 21cm absorption line(s) measurement is also a potent probe of the temperature evolution in the CMB and the intergalactic medium.  Before the light from the first stars ionizes the intergalactic gas, the neutral hydrogen resonantly absorbs the 1.42 GHz radiation line as the CMB passes through. This 1.42 GHz or 21cm-wavelength spectral line corresponds to the hyperfine energy split between aligning and anti-aligning the spin of the electron and that of the nucleus in the ground state of the neutral hydrogen, which form a spin-0 singlet and a spin-1 triplet. The population ratio between the triplet and singlet states is described by the spin temperature as $N_{1}/N_{0} = 3\ {\rm e}^{-0.068{\rm K}/T_{\rm S}}$. The 21cm absorption intensity from the radiation background, i.e. the CMB, is given by the brightness temperature~\cite{Zaldarriaga:2003du},
\be 
T_{21}\approx 0.023{\rm K}\cdot x_{\rm H_I}(z)\left(\frac{0.15}{\Omega_{\rm m}}\cdot\frac{1+z}{10}\right)^{\frac{1}{2}} 
\frac{\Omega_{\rm b} h}{0.02}\left(1-\frac{T_{\rm CMB}}{T_{\rm S}}\right),
\label{eq:T21}
\ee
where $x_{\rm H_I}$ is the neutral (${\rm H_I}$) fraction of the intergalactic hydrogen gas. For redshift $z\ge 20$ prior to reionization time, $x_{\rm H_I}\simeq 1$ in standard astrophysics.  $\Omega_{\rm m}$ and $\Omega_{\rm b}$ are the total matter and baryon fractions of the critical energy density of the Universe, and $h$ is the Hubble constant in the unit of 100 ${\rm km\ s^{-1} Mpc^{-1}}$. The latest precision measurements of these cosmological parameters are given by the {\it Planck} experiment~\cite{Ade:2015xua}. With the presence of neutral hydrogen $x_{\rm H_I}>0$ and a colder spin temperature than the radiation background, $T_{\rm S}<T_{\rm CMB}$, the absorption feature in the CMB will emerge with $T_{21}<0$. 

\begin{figure}
\includegraphics[scale=1.]{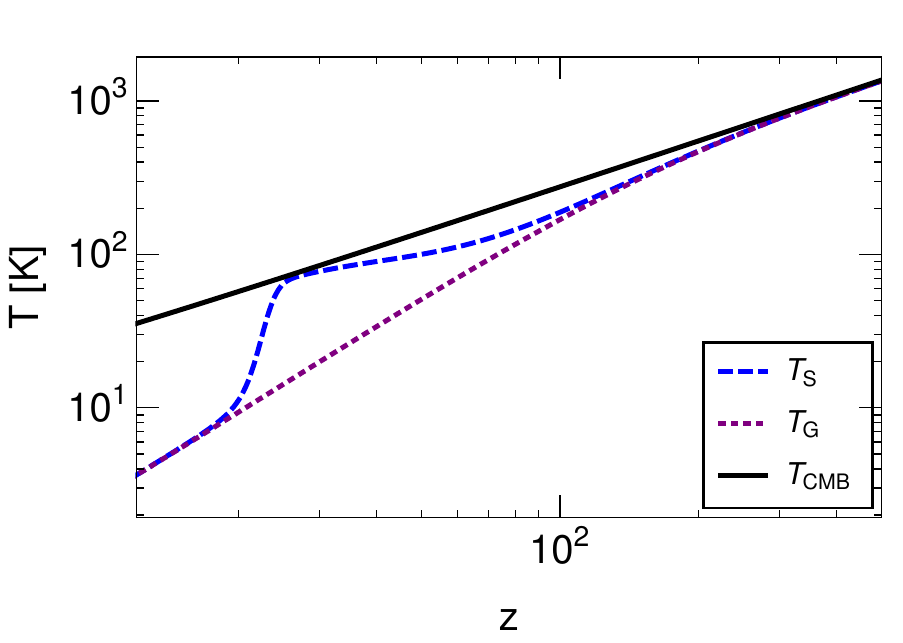}
\caption{$T_{\rm G},\ T_{\rm S}$ and $T_{\rm CMB}$ evolution in standard astrophysics without the energy injection from dark matter and black holes. $T_{\rm S}$ approaches to $T_{\rm CMB}$ after the $z\sim 200$ decoupling, suppressing the 21cm absorption until $T_{\rm S}$ recouples to $T_{\rm G}$ after the formation of first stars.}
\label{fig:standard}
\end{figure}

The hydrogen atoms decouple from the CMB at $z\sim 200$. The background radiation temperature scales with redshift as $T_{\rm CMB}=2.7{\rm K}\cdot (1+z)$, while the matter temperature scales as $(1+z)^{2}$ and cools faster than the CMB after decoupling. Hence for the hydrogen gas its $T_{\rm S}$ and $T_{\rm G}$ drop below $T_{\rm CMB}$ during the cosmic `dark age', as shown in Fig.~\ref{fig:standard}. The CMB photons can still flip the ${\rm H_I}$ hyperfine states and slowly bring $T_{\rm S}$ closer to $T_{\rm CMB}$, and in this period we typically expect $T_{{\rm G}}<T_{\rm S}<T_{\rm CMB}$. Entering the reionization epoch, the Lyman-$\alpha$ emissions from stars recouple $T_{\rm S}$ to $T_{\rm G}$ through the Wouthuysen-Field effect, and $T_{\rm S}$ demonstrates a rapid drop to the colder $T_{\rm G}$. This leads to a drop in $T_{21}$ and the expectation of a 21cm absorption signal. The EDGES measured a rapid lowering of $T_{\rm S}$ at $z\simeq 21$, that would require $T_{\rm S}\simeq T_{\rm CMB}$ to be reached before $z\simeq 21$, and $T_{\rm S}$ quickly re-couples to $T_{\rm G}$ by $z\sim 17-18$~\cite{Bowman:2018yin}.  While the timing of temperature changes in EDGES data agree with the standard evolution history~\cite{Cohen:2016jbh}, EDGES measured a $T_{\rm S}=-500$ mK~\cite{Bowman:2018yin} in the redshift range $z=15-20$, which is more than twice compared to the expectation from standard astrophysics. The flat shape of $T_{21}(z)$ in this redshift range is also unaccounted for in a standard evolution process~\citep{Bowman:2018yin}. Ref.~\cite{Barkana:2018lgd} reported this low $T_{21}$ result as a 3.8$\sigma$-strong absorption excess, and that the widened gap between $T_{\rm S}$ from $T_{\rm CMB}$ may rise from new physics.

However, for radio astronomy observations, the foreground contamination is at least four orders of magnitude higher than the 21-cm brightness temperature. This makes it extremely tricky to remove and measure the underlying signal. In Ref.~\cite{Bowman:2018yin}, the EDGES group used the polynomial foreground model to fit the Galactic synchrotron and atmospheric signal in frequency space and remove it. But there could be some low-level foreground or systematics in the system that can potentially bias the results. The EDGES data can be tested and verified by future 21cm experiments like PRIZM~\cite{Peterson:2014rga}, HERA~\cite{DeBoer:2016tnn}, LEDA~\cite{2017arXiv170909313P}, and SKA~\cite{Pritchard:2015fia}. Here, we adopt a similar approach as in Ref.~\cite{DAmico:2018sxd}, that such detection of strong 21cm absorption by EDGES would constrain the amount of accumulated high-energy particle injection that could have heated up $T_{\rm G}$ by the reionization epoch, which would narrow down the difference between $T_{\rm S}$ and $T_{\rm{CMB}}$ for $z = 15-20$ and cause significant reduction in the 21cm absorption signal. We will explore this constraint in light of steady electron and/or photon injection from decaying particle dark matter and evaporating black holes, and compare their lifetime bounds to other current limits.

\section{Energy injection effects}
\label{sect:injection}

Decaying dark matter particles with a lifetime much longer than the age of the Universe can be a steady source of the Standard Model (SM) particles. The stable particles from such injection, the photons, electron/positron and to a generally low fraction of (anti)protons can collide with and deposit energy to the intergalactic medium. The main effects from such energy deposition include enhanced ionization of the hydrogen, leading to corrections in $x_{\rm e}, x_{\rm H_I}$, and higher gas temperature $T_{\rm G}$, especially at low redshift as the energy injection can build up over time. A higher ionization fraction $x_{\rm e}$ leads to earlier reionization and more damping in the CMB's temperature and polarization correlation spectra, see Ref.~\cite{Slatyer:2016qyl, Slatyer:2015jla, Slatyer:2015kla, Clark:2016nst} for recent studies with the {\it Planck} data. For 21cm measurements, both the corrections to $x_{\rm e}$ and $T_{\rm G}$ can affect $T_{21}$, especially at a time when $T_{\rm S}$ re-coupled to $T_{\rm G}$. A reasonable choice is at the central redshift $z\simeq 17$ where EDGES detected absorption signals. By requiring the heating from new physics raises the radiation temperature by $\Delta T_{21}$ no more than $100$ or $150$ mK, this limit corresponds to a less than half or $3/4$ suppression of the standard astrophysical $T_{21}=-200$ mK absorption strength. In standard astrophysics this temperature rise can wipe out or greatly suppress the 21cm absorption signal. It is also larger than EDGES's $T_{21}$ $1\sigma$ up-fluctuation uncertainty ($+200$ mK by $99\%$ credence level~\cite{Bowman:2018yin}). 

\subsection{Decaying dark matter}

The decay of dark matter is insensitive to the small-scale matter density distribution and gives a steady energy injection rate,
\be 
\frac{{\rm d}E}{{\rm d}V{\rm d}t}= \Gamma_{\rm DM}\cdot \rho_{{\rm c,0}}\Omega_{\rm DM} (1+z)^3,
\label{eq:inj_decay}
\ee
where $\Gamma$ is the dark matter decay width, $\rho_{{\rm c,0}}$ is the current critical density of the Universe. In comparison to the $(1+z)^6$ redshift dependence in the DM annihilation case, the injection rate from DM decay drops much slower than that in annihilation, and can be more significant at lower $z$.

The photons and electrons are injected at high energy that can typically reach up to $O(10^{-1})M_{\rm DM}$. They gradually lose energy by interacting~\cite{Slatyer:2012yq, Belotsky:2014twa, Liu:2016cnk} with the intergalactic medium via ionization, Lyman-$\alpha$ excitations, gas temperature heating, as well as scattering off the background continuum photons that is studied in Ref.~\cite{Yang:2018gjd} as another explanation of the EDGES data with a heated photon radiation background. Being relativistic, these particles may take a long time to deposit all their energy into the environment. Each energy deposition channel's rate will accumulate contribution from all injection from earlier times. 

The energy deposition introduces additional terms in the evolution of ionization fraction and the Hydrogen temperature:
\bea 
\frac{{\rm d}x_{\rm e}}{{\rm d}z} & = & \left.\frac{{\rm d}x_{\rm e}}{{\rm d}z}\right|_{\rm orig} - \frac{1}{(1+z)H(z)}[I_{\rm X_i}(z)+I_{\rm X_\alpha}(z)], \label{eq:dxedz_full}\\
\frac{{\rm d}T_{\rm G}}{{\rm d}z} & = & \left.\frac{{\rm d}T_{\rm G}}{{\rm d}z}\right|_{\rm orig} - \frac{2}{3k_{\rm B}(1+z)H(z)}\frac{K_{\rm h}}{1+f_{\rm He}+x_{\rm e}}.
\label{eq:dTGdz_full}
\eea
In the additional terms, $f_{\rm He}$ is the Helium fraction in the intergalactic medium, $k_{\rm B}$ and $H(z)$ are the usual Boltzmann constant and the Hubble parameter. The $I_{\rm X_i}~ (I_{\rm X_\alpha})$ factors correspond to the energy deposition into ionization from the Hydrogen ground (excited) states. $K_{\rm h}$ takes account of the heating of intergalactic gas. These factors relate to energy injection rate by
\bea
I_{\rm X_i}(z) & = & \frac{f_{\rm i}(E,z)}{{\rm H}_H(z) E_{\rm i}}\frac{{\rm d}E}{{\rm d}V{\rm d}t},\label{eq:Ii}\\
I_{\rm X_\alpha}(z) & = & (1-C)\frac{f_\alpha(E,z)}{n_{\rm H}(z) E_\alpha}\frac{{\rm d}E}{{\rm d}V{\rm d}t}, \label{eq:Ia}\\
K_{\rm h}(z) & = & \frac{f_{\rm h}(E,z)}{n_{\rm H}(z)}\frac{{\rm d}E}{{\rm d}V{\rm d}t}\label{eq:Kh}\\
C & = & \frac{1+K \Lambda_{2s,1s} n_{\rm H} (1+x_{\rm e})}{1+K \Lambda_{2s,1s} n_{\rm H} (1-x_{\rm e})+K \beta_{\rm B} n_{\rm H} (1-x_{\rm e})}.\label{eq:C_rate}
\eea
$n_{\rm H}$ is the Hydrogen number density, and $E_{\rm i}, E_\alpha$ are the electron energy levels at the ground and excited states of the Hydrogen atom. $\Lambda_{2s,1s}$ is the decay rate from the $2s$ to $1s$ energy level. $\beta_{\rm B}$ is the effective photoionization rate, and $K=\lambda_\alpha^3/(8\pi H(z)$, with $\lambda_\alpha$ as the Lyman-alpha wavelength. $C$ is approximately the probability of an excited Hydrogen atom to emit a photon before becoming ionized~\cite{Madhavacheril:2013cna}. The redshift dependence effective efficiencies $f_{\rm i},f_\alpha, f_{\rm h}$ represent the ratio of energy deposition into each channel to the total energy injection rate at the current redshift. These effective efficiencies $f(E,z)$ will include both species/spectrum averaging and the accumulative contribution from earlier injections, after propagating to the current redshift. Thus $f(E,z)$ has dependence on the injection history, the cosmic ray species and their energy at injection. 

We calculate the effective efficiencies for different models based on the numerical values given in Refs.~\cite{Liu:2016cnk, Slatyer:2015kla} for electron and photon effective efficiency maps. These are the only maps required as only electrons and photons can efficiently deposit energy into the intergalactic medium. Protons may also be produced if kinematically allowed, but its heating contribution can be ignored due to much sub-leading final state multiplicity. Efficiency maps for other products are created by first calculating model dependent immediate decay products. These products then undergo a decay chain that produces spectra of stable products. This decay is assumed to occur instantaneously. Finally, the spectra are combined with the original electron and photon efficiency tables through a weighted average to create the effective efficiency map for the model~\cite{Clark:2017fum}
\begin{equation}
f_c(m_{\rm DM},z) = \frac{\sum_s \int f_c(E,z,s) E ({\rm d}N/{\rm d}E)_s {\rm d}E}{\sum_s \int E ({\rm d}N/{\rm d}E)_s {\rm d}E},
\label{equ:eff_calc}
\end{equation}
where $c$ is the channel, $s$ is the species, $f_c(E,z,s)$ is the effective efficiency for the channel and species, and $(dN/dE)_s$ is the spectrum for the species.

Equations~\ref{eq:Ii}$-$\ref{eq:Kh} take in the energy deposit corrections on Hydrogen ionization, Lyman-$\alpha$ excitation and gas heating.  We do not include the effects on Helium ionization and the energy loss to the photon continuum as their impact is subdominant~\cite{Clark:2016nst}.

The terms with lower script `orig' in Eqs.~\ref{eq:dxedz_full} and ~\ref{eq:dTGdz_full} refer to the unaltered standard evolution equations~\cite{Madhavacheril:2013cna,AliHaimoud:2010dx},
\bea
\left.\frac{{\rm d}x_{\rm e}}{{\rm d}z}\right|_{\rm orig} &=& \frac{C}{(1+z)H(z)}\times(x_{\rm e}^2 n_{\rm H}\alpha_{\rm B}-\beta_{\rm B} (1-x_{\rm e}) e^{-h \nu_{2s}/k_{\rm B} T_{\rm G}})
, \label{eq:dxedz_orig}\\
\left.\frac{{\rm d}T_{\rm G}}{{\rm d}z}\right|_{\rm orig} &=& \frac{8 \sigma_{\rm T} a_{\rm R} T_{\rm{CMB}}^4}{3 m_{\rm e} c H(z) (1+z)}\frac{x_{\rm e}}{1+f_{\rm{He}}+x_{\rm e}}(T_{\rm G}-T_{\rm{CMB}})
,
\label{eq:dTGdz_orig}
\eea
where $\alpha_{\rm B}$ is the effective recombination.

\begin{figure}
\includegraphics[scale=0.8]{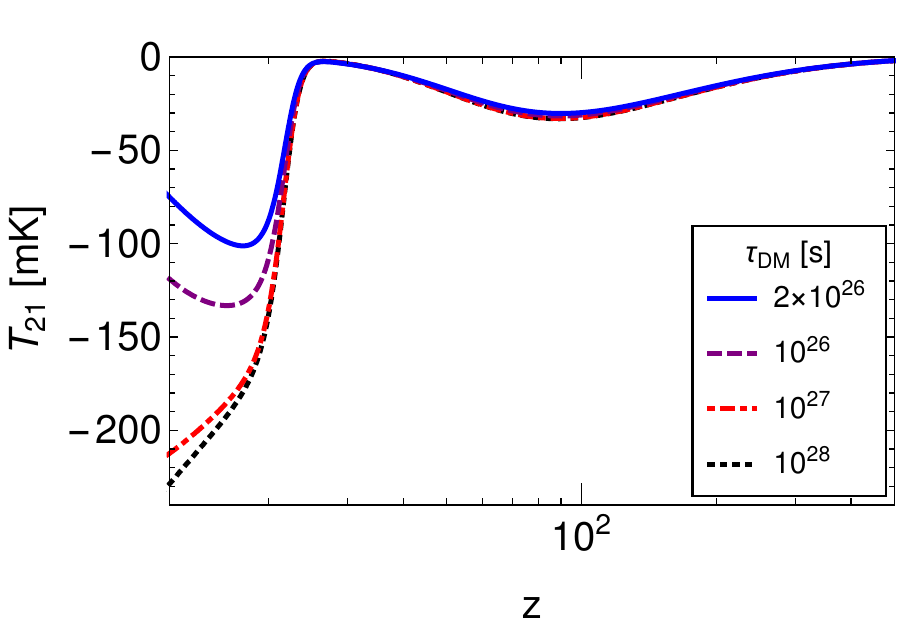}
\includegraphics[scale=0.8]{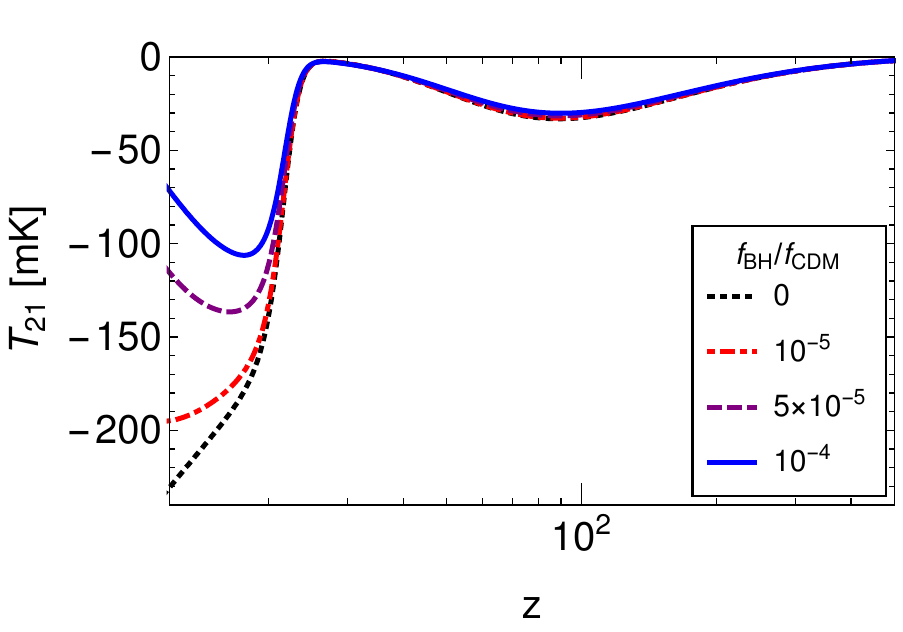}
\caption{Dark matter decay (left) and primordial black hole evaporation (right) effects lead to higher $T_{21}$ in the reionization epoch. Here dark matter mass has 100 GeV mass and decays into an $e^+e^-$ final state. The black mass is assumed to $10^{16}$ g.}
\label{fig:benchmarks}
\end{figure}

We use the numerical package {\bf HyRec}~\cite{AliHaimoud:2010dx} to compute the temperature evolutions, with the energy injection corrections implemented into the evolution equations. The Wouthuysen-Field effect is included into the calculation by defining \cite{Zaldarriaga:2003du}
\bea
T_{\rm S} & = & \frac{T_{\rm{CMB}}+y_{\rm c} T_{\rm G}+y_{\rm{Ly\alpha}}T_{\rm{Ly\alpha}}}{1+y_{\rm c}+y_{\rm{Ly\alpha}}}, \label{eq:T_s}\\
y_{\rm c} & = & \frac{C_{10}}{A_{10}} \frac{T_\star}{T_{\rm G}}, \label{eq:y_c}\\
y_{\rm{Ly\alpha}} & = & \frac{P_{10}}{A_{10}} \frac{T_\star}{T_{\rm{Ly\alpha}}}, \label{eq:y_Lya}
\eea
where $A_{10}=2.85\times10^{-15}s^{-1}$ is transition's spontaneous emission coefficient, $C_{10}$ is the collisional de-excitation rate of the triplet hyperfine level, $P_{10}\approx 1.3\times 10^{-12}J_{-21}s^{-1}$ is the indirect de-excitation rate due to Lyman-Alpha absorption, $T_\star=h\nu_0/k_{\rm B}=0.068$ K is the Lyman-Alpha energy, $T_{\rm{Ly\alpha}}$ is the Lyman-Alpha background temperature, and $T_{\rm{Ly\alpha}}=T_{\rm G}$ for the period of interest, and $J_{-21}$ is the Lyman-Alpha background intensity and is estimated by the early and late reionization results of \cite{Ciardi:2003hg}. We use the cosmological parameters $\Omega_{\rm m}=0.3$, $\Omega_{\rm b}=0.04$, $\Omega_\Lambda=0.7$, and $h=0.7$ throughout this work. 
Fig.~\ref{fig:benchmarks} illustrates the heating effect on $T_{21}$ from dark matter decay, assuming contribution from 100\% of the relic density and the DM$\rightarrow e^+e^-$ channel. Heating of neutral Hydrogen becomes manifest at near-reionization time.
Note that variations in cosmological parameters do slightly affect the result. Cosmological parameter variation within {\it Planck}'s constraint is expected to lead to ${\cal O}(1)$ correction. As an example, for the best fit of {\it Planck}'s TT,TE,EE+lowP data, $\Omega_{\rm m}=0.316$, $\Omega_{\rm b}=0.049$, $\Omega_\Lambda=0.684$, and $h=0.67$~\cite{Ade:2015xua}, the constraints shown in Section~\ref{sect:constraints} weaken by a factor of 1.3. 

\subsection{Primordial black holes}

Another potential source of steady $e^\pm,\gamma$ injection is the Hawking radiation of long-lived, relatively low mass ($M_{\rm BH}>10^{15}$g) primordial black holes. Over-density in the early Universe can collapse into primordial black holes~\cite{Hawking:1971ei, pbh2, pbh3, Carr:1974nx}. If PBH formation occurs during radiation dominated phases, usually a horizon-sized fluctuation is needed to overcome the radiation pressure and makes over-density growth possible, leading to a characteristic PBH size. In matter dominated phases, however, PBH formation can be a lot more complicated. The lack of radiation pressure allows for black hole formation over a wide range of mass, and the mass profile can depend on the geometric symmetry of density fluctuations~\cite{Khlopov:2008qy, nonthermal2}. For a review of PBH formation and relevant constraints, see Ref.~\cite{Carr:2016drx,Green:2016xgy}. Also see Ref.~\cite{Georg:2016yxa,Chen:2016pud,Georg:2017mqk} for recent studies of PBH formation under nonthermal conditions. 

Relevant PBHs for post-recombination energy injection need to be long-lived such that its evaporation time scale is longer than the age of Universe. PBHs with $M_{\rm BH}> 10^{15}$g can survive to today, and the PBH in the mass range $10^{15}-10^{17}$g are subject to indirect searches of extragalactic cosmic rays~\cite{Carr:2009jm} and CMB damping constraint~\cite{Clark:2016nst}.

A black hole of mass $M_{\rm BH}$ gives away its mass at the Hawking radiation rate~\cite{Hawking:1974rv},
\be 
\dot{M}_{\rm BH} = -5.34\times 10^{25} \left(
\sum_{i} \phi_i
\right) M_{\rm BH}^{-2}\ \ {\rm g^3s^{-1}},
\label{eq:hawking}
\ee
where the coefficients $\phi_i$ is the fraction of evaporation power and sums over all particle degrees of freedom that are lighter in mass than the BH's temperature $T_{\rm BH}= (8\pi G M_{\rm BH})^{-1}$. Here we use the Greek letter $\phi$ to avoid confusion with the effective absorption coefficient $f_{\rm i}$. The relevant emission are photons and electrons as they can interact with the intergalactic medium. Other emission species, like neutrinos, do not deposit their energy into the intergalactic medium in an efficient manner. For each particle degree of freedom in photons and electrons, $\phi^\gamma_1= 0.06$ and $\phi_{1/2}^{e^{\pm}}=0.142$~\cite{MacGibbon:1991tj}. Note these $\phi$ values are normalized to the emission of a $10^{17}$g black hole. The PBH injection also scales as $(1+z)^3$ and depends on the abundance of black holes,
\be 
\frac{{\rm d}E}{{\rm d}V{\rm d}t} = \sum_{i=\gamma,e^\pm}\phi_i\cdot \frac{\dot{M}_{\rm BH}}{M_{\rm BH}}\rho_{\rm c,0}\Omega_{\rm BH} (1+z)^3,
\label{eq:inj_pbh}
\ee
where $\dot{M}_{\rm BH}/M_{\rm BH}\propto M_{\rm BH}^{-3}$ is a mass loss rate. For $M_{\rm BH}\gg 10^{15}$g, $M_{\rm BH}$ can be consider within the age of the Universe. Comparing with Eq.~\ref{eq:inj_decay}, PBH's injection rate has the same redshift dependence as that in the dark matter decay scenario.
The treatment of interaction of photon and electrons with the intergalactic medium follows the same procedure as discussed in the previous subsection. The impact on $T_{21}$ is shown in the right panel of Fig.~\ref{fig:benchmarks} for a $10^{16}$g mass PBH with a few sample abundance between $10^{-5}$ and $10^{-4}$ of the Universe's matter density.

\section{Constraints from 21cm}
\label{sect:constraints}

By requiring the $T_{21}$ correction to its standard astrophysical value at $z\simeq 17$ to be less than 100 and 150 mK, namely $T_{21}(z=17)<-100$ and $-50$ mK respectively, we obtain strong constraints on the lifetime of decaying dark matter, and the maximally allowed abundance of primordial black holes.

\begin{figure}[h]
\includegraphics[scale=0.9]{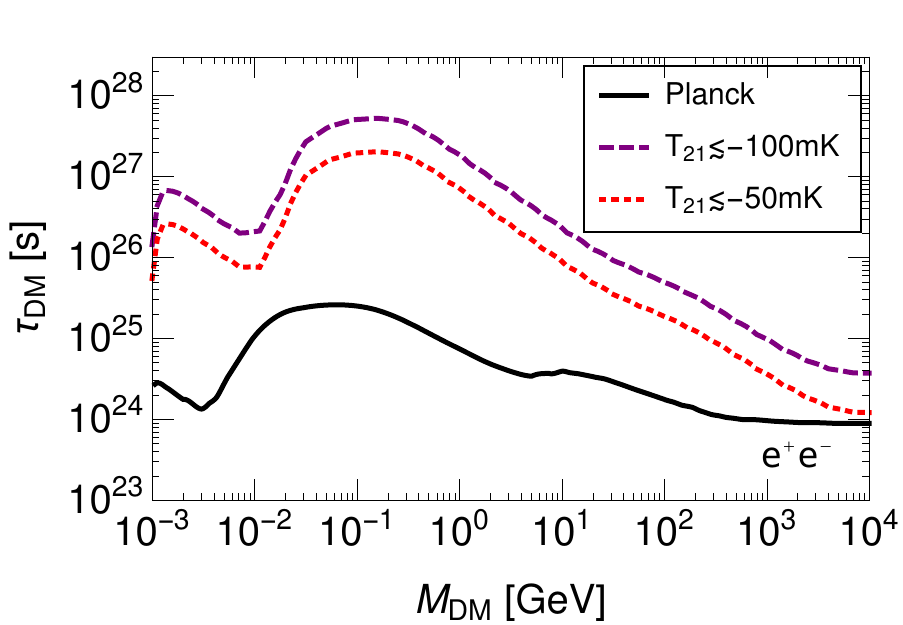}
\includegraphics[scale=0.9]{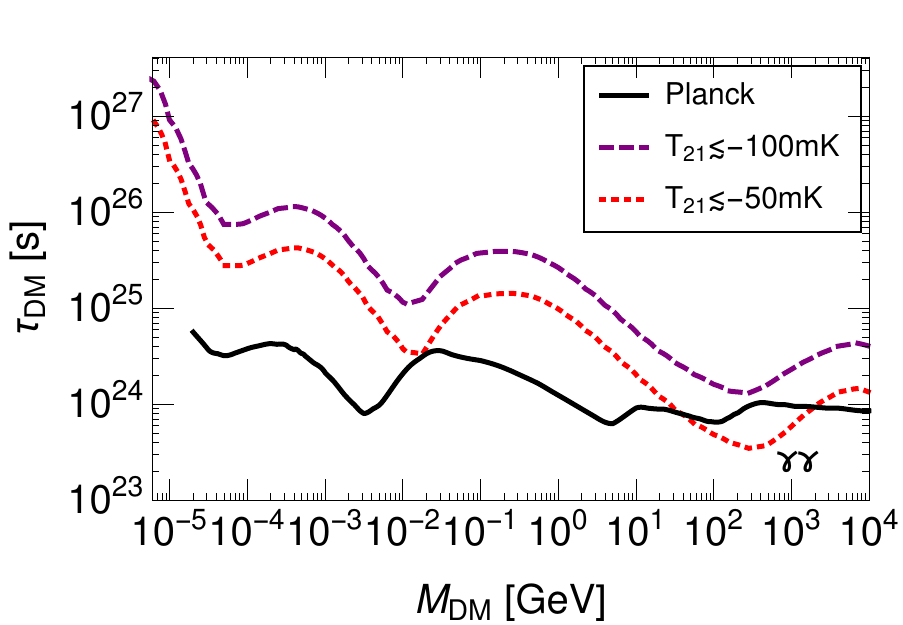}
\includegraphics[scale=0.9]{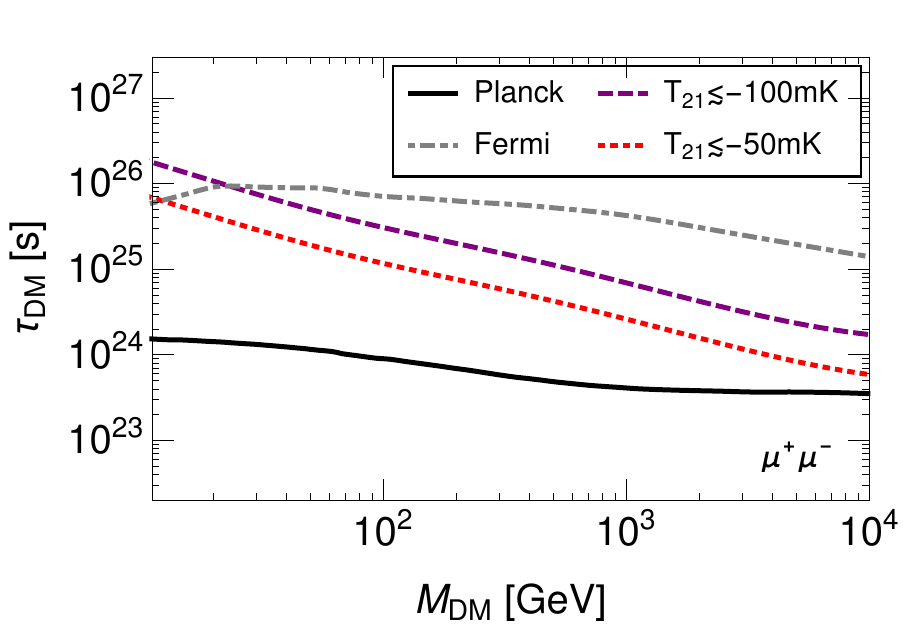}
\includegraphics[scale=0.9]{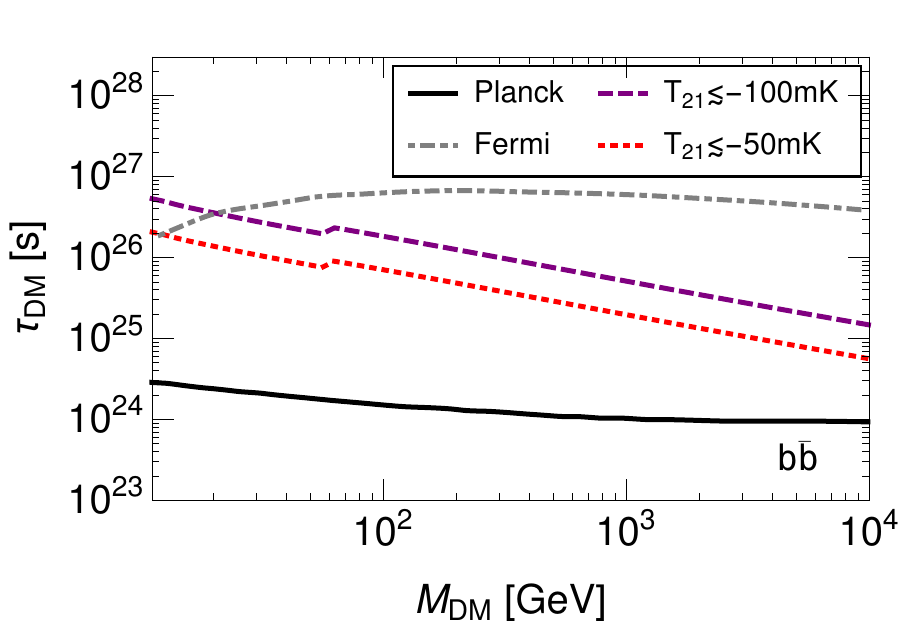}
\caption{21cm {\it lower}-bounds on dark matter decay lifetime, and primordial black hole abundance. The DM decay panels assume ${\rm DM}\rightarrow e^+e^-$ (top left), ${\rm DM}\rightarrow \gamma\gamma$ (top right), ${\rm DM}\rightarrow \mu^+\mu^-$ (bottom left) and ${\rm DM}\rightarrow b\bar{b}$ (bottom right) final states. Current CMB damping constraints~\cite{Slatyer:2016qyl} from {\it Planck} (solid) and dwarf galaxy bounds~\cite{Baring:2015sza} from Fermi-LAT (gray dashed) are also shown for comparison.}
\label{fig:bounds}
\end{figure}

Fig.~\ref{fig:bounds} illustrates the constraint on the decay lifetime $\tau_{\rm DM}$ for DM mass from MeV up to 100 TeV. The constraint assumes generic two-body decay channels. The DM$\rightarrow e^+e^-$ channel is the most stringently constrained due to its highest fraction of electrons in the final state. $\mu^+\mu^-$ and $b\bar{b}$ final states are also plotted, which have lower $f(E,z)$ in comparison. $\mu^+\mu^-, b\bar{b}$ are also much smoother than $e^+e^-$ due to the wide spectra of stable final particles which results in most features of $f$ averaging out. As lower energy injection requires less time to deposit its energy into the intergalactic medium, $f$ increases with lower $M_{\rm DM}$, as demonstrated in the shape of $\tau_{\rm DM}$ constraint. This leads to a significant ${\cal O}(10^{27})$s bound for sub-GeV dark matter lifetime that is complementary to gamma ray search limits~\cite{Baring:2015sza, Liu:2016ngs} from Fermi-LAT data. The 21cm bound is also stronger than the CMB damping constraint from {\it Planck}~\cite{Clark:2016nst} by more than one order of magnitude. This indicates that the $T_{\rm S}\simeq T_{\rm G}$ in the reionization epoch is also a very sensitive test of energy injection. $\mu^+\mu^-$ and $b\bar{b}$ final states produce weaker bounds than gamma-ray measurements from Fermi-LAT~\cite{Baring:2015sza} for masses above 20 GeV. However, because $e^+e^-$ produces fewer gamma-rays and has a higher $f(E,z)$, it is expected to be much more constraining than Fermi-LAT. The shape of the constraints is a direct result from the effective efficiency maps discussed in Sec.~\ref{sect:injection}. Masses that occur near a peak absorption efficiency have a corresponding high constraint. The shifting of the peaks between Planck and 21cm results is due to the $z$ dependence of the effective efficiency. Dominant features present in the efficiency map shift to higher DM and lower PBH masses at late redshift and are observed in calculated maps~\cite{Clark:2016nst,Liu:2016cnk}.

Also note the enhanced lifetime in the $\gamma\gamma$ channel at injection below 0.1 MeV due to higher photon energy absorption efficiency. At $\sim$KeV mass DM the lifetime bound is higher than $10^{27}$s. This bound is below the $10^{29}$s$\cdot (M_{\rm DM}/$KeV$)$~\cite{Bulbul:2014sua} requirement for explaining the 3.5 KeV X-ray excess~\cite{Boyarsky:2014jta}. Testing this signal would need ${\cal O}({\rm mK})$ $T_{21}$ sensitivity at future measurements.

\begin{figure}
\includegraphics[scale=1.2]{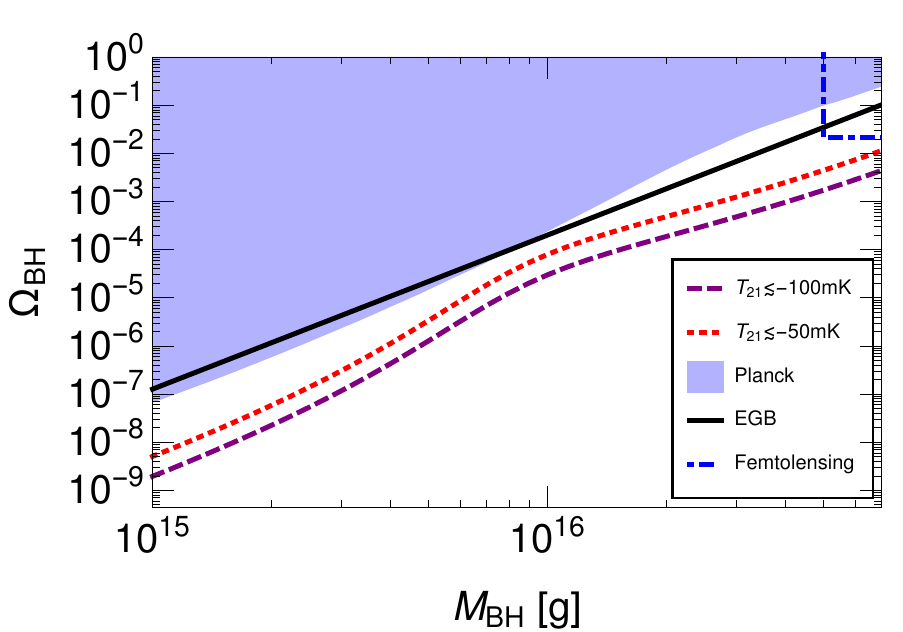}
\caption{21cm {\it upper}-bounds on the primordial black hole abundance. Current CMB damping~\cite{Poulin:2016anj,Clark:2016nst} (shaded), extragalactic gamma-ray background~\cite{Carr:2016drx} (solid) constraints and the femtolensing excluded area~\cite{Barnacka:2012bm} (top right) are also shown for comparison.}
\label{fig:PBH_bound}
\end{figure}

For PBHs, the injection rate $\dot{M}_{\rm BH}/{M_{\rm BH}}\propto {M_{\rm BH}}^{-3}$ quickly drops for higher BH masses. Also, for BH mass much higher than $10^{16}$ g, the BH temperature drops below the electron mass, reducing the amount of electron injection and the impact on the intergalactic medium's temperature. Fig.~\ref{fig:PBH_bound} shows the 21cm-constrained maximal fraction of the Universe's dark matter in the form of primordial black holes. Comparing with CMB damping limit from {\it Planck}, the 21cm bound is stronger by one order of magnitude throughout the relevant mass range.

\bigskip

\section{Conclusion}

We studied the dark matter decay and PBH radiation's impact on the 21cm radiation temperature in the reionization epoch, in light of recent measurement of 21cm absorption signal by EDGES experiment. In this work we do not aim at resolving the deviation of EDGES's measured $T_{21}$ from the standard astrophysical expectation. Instead we impose a constraint on the historic energy injection in the intergalactic medium, by requiring this injection not wipe out or cause major correction to the 21cm absorption signal by $z=17$, at the middle of redshift range where the absorption signal is expected to be the strongest. 

We adopt the effective absorption efficiencies from Ref.~\cite{Slatyer:2015kla} for a continuous energy deposit process from injected particles, and simulate the spin and gas temperatures down to redshift $z=17$. We considered DM decay channels $DM\rightarrow e^+e^-, \gamma\gamma, \mu^+\mu^-, b\bar{b}$ and obtained  $\tau_{\rm DM}\ge 10^{26-27}$s bounds on the DM lifetime by requiring the heating process raises the gas temperature to no higher than -100 mK or -50 mK. For $e^+e^-$, $\gamma\gamma$ final states  and PBH cases, the 21cm observation provides the best bound in the DM mass-lifetime parameter space. For $b\bar{b}$ and $ \mu^+\mu^-$ final states, the 21cm observation bound becomes better than all the existing constraint for $m_{\rm {DM}}<20$ GeV. In both DM and PBH cases, the 21cm bound is found to be better than current CMB damping constraint from {\it Planck} data.

Since the removal of extremely large foreground from the data  is difficult, the EDGES result needs to be verified by other experiments. The future 21cm experiments like PRIZM, HERA, LEDA, and SKA will be able to verify EDGES results. If the absorption signal is verified in the future, the 21cm absorption can prove to be a powerful probe to non-standard heating processes. It is worth emphasizing on the 21cm's sensitivity to $e^\pm,\gamma$ injection in the sub-GeV energy range as demonstrated in Fig.~\ref{fig:bounds}. In contrast to the poor absorption efficiency at TeV or higher energy scale, the sub-GeV bound on decaying dark matter can nicely fill in the MeV-GeV range where the indirect search bounds are current less stringent in comparison to X-ray and hard gamma ray limits.

{\bf Acknowledgements}

SJC acknowledges support from NASA Astrophysics Theory grant NNX12AC71G. BD and LES acknowledge support from DOE Grant de-sc0010813. Y.G. is supported under grant no.~Y7515560U1 by the Institute of High Energy Physics, Chinese Academy of Sciences. Y.Z.M is supported by the National Research Foundation of South Africa with Grant no.~105925 and no.~104800. We thank X-J.Bi, N. Mirabolfathi, N. Suntzeff for discussions.

\bibliography{refs}

\end{document}